\documentclass[11pt]{article}

\usepackage{mathptmx}
\usepackage{caption}

\usepackage{amsmath,amssymb,amsfonts,amsthm} 
\usepackage{mathrsfs,latexsym} 
\usepackage{bm}
\usepackage{calrsfs}
\DeclareMathAlphabet{\pazocal}{OMS}{zplm}{m}{n}

\newcommand{\RR}{{\mathbb R}}

\newcommand{\Dc}{{\mathcal{D}}}

\newcommand{\bai}{{\bm a}}
\newcommand{\bbi}{{\bm b}}

\newcommand{\bfi}{{\bm f}}
\newcommand{\bgi}{{\bm g}}
\newcommand{\bxi}{{\bm x}}

\newcommand{\bqi}{{\bm q}}

\newcommand{\bPi}{{\bm P}}
\newcommand{\bQi}{{\bm Q}}

\newcommand{\Alg}{\pazocal{A}}                   
\newcommand{\Group}{\pazocal{G}}
\newcommand{\Lag}{\pazocal{L}}              

\newcommand{\ie}{\textit{i.e.\ }}
\newcommand{\eg}{\textit{e.g.\ }}

\begin{document}
\title{From path integrals to dynamical algebras: \\ a macroscopic 
view of quantum physics}

\author{\large Detlev Buchholz${}^{(1)}$ \ and 
\ Klaus Fredenhagen${}^{(2)}$ \\[5mm]
\small 
${}^{(1)}$ Mathematisches Institut, Universit\"at G\"ottingen, \\
\small Bunsenstr.\ 3-5, 37073 G\"ottingen, Germany\\[5pt]
\small
${}^{(2)}$ 
II. Institut f\"ur Theoretische Physik, Universit\"at Hamburg \\
\small Luruper Chaussee 149, 22761 Hamburg, Germany \\
}
\date{}

\maketitle

\begin{abstract}
The essence of the path integral method in quantum physics can be expressed
in terms of two relations between unitary propagators,  
describing perturbations of the underlying system. They inherit the 
causal structure of the theory and its invariance properties under 
variations of the action. These relations determine a dynamical algebra 
of bounded operators which encodes all properties of the corresponding 
quantum theory. This novel approach is  
applied to non-relativistic particles, where  
quantum mechanics emerges from it. The method works also in 
interacting quantum field theories  
and sheds new light on the foundations of quantum physics.
\end{abstract}

\section{Introduction}

Path integrals \cite{Di,FeHi} 
are a standard tool in the theoretical description
of quantum systems. Starting from a classical theory, describing paths 
(orbits) of the underlying system in configuration space and an action, 
governing its dynamics, they provide formulas for the propagators,
\ie the time ordered scattering operators in the resulting quantum 
theory. These formulas yield useful algorithms for the treatment of 
concrete problems. The proper definition of the underlying functional integrals
is a quite subtle matter, however. 
It frequently requires reformulations of the integrals, \eg in Euclidean 
space or on discrete lattices, which defy a direct physical interpretation. 
Moreover, changes of the states of the system, such as the passage from 
vacuum to thermal states, require adequate modifications of the integrals. 
And, last but not least, even though this approach is known to give
satisfactory results, its conceptual foundations remained obscure.

It is therefore gratifying that the essence of the path integral formalism can 
be replaced by simple algebraic relations without having to dive into the 
mathematical subtleties of functional integrals on infinite 
dimensional configuration spaces. Moreover, this reformulation sheds new
light on the foundations of  quantum physics. This new algebraic framework 
was established in quantum field theory in an effort to amend the 
axiomatic framework by some concrete dynamical input \cite{BuFr1}.
It was then extended to quantum mechanics, where its conceptual
foundations were also settled \cite{BuFr2}. It is the aim of the
present letter to clarify its relation to the path integral formalism.

\section{Path integrals}
Consider a system of $N$ distinguishable particles of equal 
mass (put equal to $1$)  whose orbits are 
given by functions of time $t \mapsto \bxi(t) \in \RR^{3N}$. 
Their unperturbed motion is fixed by the Lagrangean
$\Lag_0(\dot{\bxi}) = (1/2) \, \dot{\bxi}^2$. 
Temporary perturbations of this motion and interactions between the 
particles are described by time dependent potentials 
\[
\bxi \mapsto V_t(\bxi) = \sum_k g_k(t) \, V_k(\bxi) \, ,
\]
where the functions $g_k$ indicate when and for how long the 
potentials $V_k$ are effective. Their impact on any given 
orbit $\bxi$ is given by functionals
$F[\bxi] = \int_{t_i}^{t_f} \! dt \, V_t(\bxi(t))$,
where $t_i$ is any time before the perturbation starts and 
$t_f$ any time after it has finished; the square bracket $[\bxi]$ indicates
that a quantity depends on the whole orbit $\bxi$.
Turning to the quantum system, the effect of the perturbations 
can be described by a unitary scattering matrix $S(F)$. It is 
given by the formula
\[
S(F) = e^{it_f H_0} U_F(t_f,t_i) e^{-it_i H_0} \, ,
\]
where $H_0$ is the Hamiltonian of the unperturbed system 
and $U_F(t_f,t_i)$ is the propagator for the perturbed 
Hamiltonian $H_0 + V_t(\bQi)$, $\bQi$ being the position operator. 
The kernel of the propagator is given  
by a path integral in position space, involving the above 
functional $F$, 
\[
\langle \bqi | \, U_F(t_f,t_i) \, | \bqi' \rangle 
= \int_\bqi^{\bqi'} \! \Dc \bxi \, 
\exp \Big( i \int_{t_i}^{t_f} \! dt \, 
(\Lag_0(\dot{\bxi}(t)) - V_t(\bxi(t)) \Big) \, .
\] 
Ignoring mathematical subtleties,  
$\Dc \bxi$ denotes the measure on the family of orbits (paths) $\bxi$, 
starting at time $t_i$ at $\bqi$ and ending at 
time $t_f$ at $\bqi'$. Still at a heuristic level, let us assume 
that this measure is invariant under deformations of the
orbits by loops $t \mapsto \bxi_0(t)$ about $0$,
\ie smooth functions having their supports in the time 
interval $t_i \leq t \leq t_f$. The integrand is thereby transformed into 
\[
\Lag_0(\dot{\bxi}(t) + \dot{\bxi}_0(t)) - 
V_t(\bxi(t) + {\bxi}_0(t)) 
= \Lag_0(\dot{\bxi}(t)) + \delta \Lag_0(\dot\bxi_0)(\dot{\bxi}(t)) 
- V_t^{\bxi_0}(\bxi(t)) \, ,
\]
where we make use of the short hand notation
\begin{align*}
\delta \Lag_0(\dot\bxi_0)(\dot{\bxi}) & : = 
\Lag_0(\dot{\bxi} + \dot{\bxi}_0) - \Lag_0(\dot{\bxi}) 
= \dot{\bxi_0} \, \dot{\bxi} + (1/2) \, \dot{\bxi}_0^2 \\
V_t^{\bxi_0}(\bxi) & := V_t(\bxi + \bxi_0)  \, .
\end{align*} 
Now by a partial integration, the velocity $\dot{\bxi}$ in the integral
$\int_{t_i}^{t_f} \! dt \, \delta \Lag_0(\dot\bxi_0)(\dot{\bxi}(t))$
can be transformed into $\bxi$, so the term
$\int_{t_i}^{t_f} \! dt \, (\delta \Lag_0(\dot\bxi_0)(\dot{\bxi}(t)) 
- V_t^{\bxi_0}(\bxi(t))$ describes just another perturbation. 
Plugging this information into the scattering matrix for 
given perturbing functional $F$ and any given loop $\bxi_0$, and putting 
$F^{\bxi_0}[\bxi] := F[\bxi + \bxi_0]$, we arrive at the equality 
\begin{equation} \label{e1}
S(F) = S(F^{\bxi_0} + \delta \Lag_0(\dot{\bxi}_0)) \, ,
\end{equation}
containing dynamical information. 
The second fundamental ingredient in the functional integral formalism
is the time ordering involved in the definition of the scattering operators.
A perturbing functional $F_1$ is said to lie in the future of $F_2$
if the underlying potential $t \mapsto V_{1,t}$ in $F_1$ has 
support in the future of $t \mapsto V_{2,t}$, entering in $F_2$.
The preceding discussion then implies that the resulting scattering 
matrices satisfy the equality $S(F_1) S(F_2) = S(F_1 + F_2)$. 
As a matter of fact, given any other functional $F_3$, one obtains
by splitting it into $F_3 = F_{3 +} + F_{3 -}$, such that $F_1 + F_{3 +}$
lies in the future of $F_2 + F_{3 -}$, the more refined 
equality \cite{BuFr2}
\begin{equation} \label{e2}
S(F_1 + F_2 + F_3) = S(F_1 + F_3) S(F_3)^{-1} S(F_2 + F_3) \, .
\end{equation}
The preceding two equalities is all what is needed in order to define 
dynamical algebras.

\section{Dynamical algebras} 
We leave now the mathematically subtle territory of path integrals and 
take a fresh look at the problem of describing our interventions   
into the quantum world. From a macroscopic point of view there 
are two obvious facts. First, quantum systems are tested by perturbing
them; in laboratories this happens in a systematic manner, but 
its surroundings produce such perturbations as well. We therefore propose
to take the operations, induced by perturbations,  
as primary ingredients of the theory. 
If one happens to know the particular shape of some perturbation
in space and time, being described by a functional $F$, 
the resulting operation is denoted by $S(F)$. The second ingredient
taken from the macroscopic world is the arrow of time. It enters
into the quantum world since we can firmly state whether some operation
$S(F_1)$ was performed after or 
before some other operation $S(F_2)$. Moreover, it 
is impossible to make up for missed operations in the past,
time is directed. 

Since operations can be performed time and again, 
it is meaningful to assume that they form 
a semigroup. As a matter of fact, dealing
with experimentally accessible systems, it is
also plausible that the effect of some operation can be 
rubbed out by another one, in 
accordance with the fact that experiments
can be repeated. We therefore postulate that the operations $S(F)$, 
which can be performed on quantum systems, form a group. It becomes
a dynamical group for given Lagrangean. For the case at hand
of a mechanical system, we take for simplicity the free Lagrangean $\Lag_0$
and rely on findings obtained in the preceding step.

\medskip 
\noindent \textbf{Definition:} Let $\Lag_0$ be a Lagrangean.
An operation, fixed by a functional 
$F$ describing some perturbation of the dynamics, is represented 
by the symbol $S(F)$. So, in particular, 
$S(0) = 1$. These symbols generate a dynamical group 
$\Group_{\Lag_0}$. They satisfy the relations 
\begin{itemize}
\item[(i)] 
$S(F) = S(F^{\bxi_0} + \delta \Lag_0(\dot{\bxi}_0))$ \ 
for any functional $F$~and loop $\bxi_0$
\item[(ii)] 
$S(F_1 + F_2 + F_3) = S(F_1 + F_3) S(F_3)^{-1} S(F_2 + F_3)$ \
for any functional $F_3$, provided $F_1$ lies in the future of $F_2$. 
\end{itemize}

Based only on macroscopic (classical) concepts, the 
dynamical group is defined by these relations; there is   
no quantization condition built in from the outset. Nevertheless, 
the group is inherently non-commutative. As we shall see, the value of 
Planck's constant is determined by operations 
corresponding to the constant functionals
$\bxi \mapsto F_{\textrm h}[\bxi]:= {\textrm h}$, ${\textrm h} \in \RR$. 
These functionals have empty support in time; phrased differently,
they can be realized by time integrals $\int \! dt \, h(t) = {\textrm h}$,
where $t \mapsto  h(t)$ has arbitrary support. It therefore follows
from the causality condition (ii) that 
$S(F) S(F_{\textrm h}) = S(F + F_{\textrm h}) = S(F_{\textrm h}) S(F)$.
So the operations ${\textrm h} \mapsto S(F_{\textrm h})$ form an abelian group 
in the center of $\Group_{\Lag_0}$. Without essential loss of generality,  
we fix a scale and 
put $S(F_{\textrm h}) = e^{i {\textrm h} } 1$, ${\textrm h} \in \RR$. 
We also note that the dynamical relation (i) implies 
$S(\delta \Lag_0(\dot{\bxi}_0)) = 1$ for all loops; these
equalities correspond to the classical Euler-Lagrange equation. 
 
Having defined the group $\Group_{\Lag_0}$, it is a standard procedure
to construct a corresponding dynamical algebra
$\Alg_{\Lag_0}$. One declares that the elements $S_k$ of the
group are basis elements of some complex vector space, leading to
finite sums with complex coefficients,  $\sum_k c_k S_k $. The 
adjoint operators are defined by 
$(\sum_k c_k S_k)^*:= \sum_k \overline{c}_k S_k^{-1}$, where the bar
denotes complex conjugation. The rules for multiplication are 
inherited from the group by the distributive law. 
One can show \cite{BuFr2} that the algebra $\Alg_{\Lag_0}$ is  
equipped with a norm which promotes it to a C*-algebra, \ie 
an algebra which can be realized as a norm closed subalgebra of the
algebra of bounded operators on some Hilbert space. 
 
Before showing that this algebra contains the entire 
information incorporated in the conventional quantum mechanical setting,
let us emphasize that our scheme of constructing dynamical algebras 
covers a large set of systems of physical interest. What is needed 
is that the system can be described in terms of some 
classical configuration space, on which a group acts (the loops in the 
present case), and some Lagrangean. Moreover, 
one needs a causal structure, given in the present
context by the direction of time. 
In \cite{BuFr1} this scheme was applied to an interacting scalar 
quantum field in Minkowski space for which one also has a classical
configuration space, the analogue of loops, and Lagrangeans. 
The causal structure, however, has to be replaced by the relativistic 
one, and is concretely given by the partially ordered set
of forward lightcones. With that modification, the definition
of the dynamical algebra is identical to the preceding one 
and one arrives in this manner at a consistent description 
of field theoretic models, satisfying all fundamental postulates of 
relativistic quantum field theory.

\section{Recovery of quantum mechanics}

In this section we give a brief account of results established 
in \cite{BuFr2}, showing that the dynamical algebra of mechanical 
systems entails the standard formalism of quantum mechanics. In view 
of the fact that no explicit quantization condition was incorporated 
into the dynamical algebra, we are led to the conclusion  
that the origin of quantum effects is the arrow of time, being encoded
in the operations. The specific form of commutation relations 
then follows from the Lagrangean. 
 
Proceeding from the non-interacting Lagrangean $\Lag_0$, we 
consider the simplest functionals (perturbations) of the form
$F_{\bfi}[\bxi] = \int \! dt \, \bfi(t) \bxi(t) + {\mathrm h}$.
Here $t \mapsto \bfi(t)$ is some continuous function,
the integral extends over the (compact) support of $\bfi$, 
and ${\mathrm h}$ is a constant. A convenient choice which 
simplifies subsequent formulas is 
\[
{\mathrm h}(\bfi) := 
(1/2) \iint \! ds ds' \, |s - s'| \bfi(s) \, \bfi(s') \, .
\]
Given another function $\bfi'$ for which the first two moments 
coincide with those of $\bfi$, \ie
\[
\int \! dt \, (\bfi'(t)- \bfi(t)) = 
\int \! dt \, t \, (\bfi'(t) - \bfi(t)) = 0 \, ,
\]
their difference $\ddot{\bxi}_0 = \bfi' - \bfi$ determines a  
loop function
\[
t \mapsto \bxi_0(t) = 
\int_{- \infty}^t \! ds \, (t-s) \, (\bfi' - \bfi)(s) \, .
\]
By a straightforward computation one obtains for the 
resulting functionals the equalities, cf.\ \cite{BuFr2},
\[
F_\bfi[\bxi] = F_{\bfi'}[\bxi + \bxi_0] + F_{-\ddot{\bxi}_0}[\bxi] 
= F_{\bfi'}^{\bxi_0}[\bxi] + \delta \Lag_0(\dot{\bxi}_0)[\bxi] \, . 
\]
Plugging the latter expression into the dynamical relation~(i), 
one arrives at $S(F_\bfi) = S(F_{\bfi'})$ for arbitrary pairs of 
functions $\bfi, \bfi'$ whose first two moments coincide.
In particular, $S(F_{\ddot{\bxi}_0}) = 1$ for all loop
functions $\bxi_0$. As we shall see, these relations
determine solutions of the Heisenberg equation. 

Next, let $F_\bgi[\bxi] = \int \! dt \, \bgi(t) \bxi(t) + {\mathrm h}(\bgi)$
be another functional. We want to determine the relation between the 
unitary operators $S(F_\bfi)$ and $S(F_\bgi)$. To this end we replace 
$\bfi$ as in the preceding step by a function $\bfi'$ which has support in the
future of $\bgi$. Making use of the causal
relation (ii), we obtain 
\[
S(F_\bfi) S(F_\bgi) = S(F_{\bfi'}) S(F_\bgi) = S(F_{\bfi'} + F_{\bgi}) \, .
\]
In view of the linear dependence of the 
given functionals on the underlying
orbits it is apparent that $F_{\bfi'} + F_{\bgi} - F_{\bfi' + \bgi}$
is a constant functional. By another routine computation one finds
\cite{BuFr2}
\[
F_{\bfi'}[\bxi] + F_{\bgi}[\bxi] - F_{\bfi' + \bgi}[\bxi] =
-(1/2) \langle  \bfi', \Delta \bgi \rangle \, ,
\]
where we have defined 
\[ 
\langle  \bfi', \Delta \bgi \rangle := 
\iint \! ds ds'  \bfi'(s) (s'-s) \bgi(s') \, .
\]
In view of the fact that this is a constant functional
on the space of paths,  we can proceed to 
\[
S(F_{\bfi'} + F_{\bgi}) = S( F_{\bfi' + \bgi} - 
(1/2) \langle  \bfi', \Delta \bgi \rangle) 
= S( F_{\bfi' + \bgi}) \, e^{-(i/2) \langle  \bfi', \Delta \bgi \rangle} \, ,
\] 
where  in the second equality our choice of scale for the
constant functionals entered. 
Now, by construction, the first and second
moments of the functions $(\bfi' + \bgi)$ and $(\bfi + \bgi)$ coincide,
so according to the first step we have 
$S( F_{\bfi' + \bgi}) = S( F_{\bfi + \bgi})$. By inspection of its 
defining equation, it is also clear that
$\langle  \bfi', \Delta \bgi \rangle =
\langle  \bfi, \Delta \bgi \rangle$. Thus we have arrived at 
the equality 
\[
S(F_\bfi) S(F_\bgi) = e^{- (i/2) 
\langle  \bfi, \Delta \bgi \rangle } \, S(F_{\bfi + \bgi}) \, .
\]
This relation shows that the unitary operators $S(F_\bfi)$ do not
commute amongst each other. They form a familiar Lie group, the Weyl 
group. We proceed from this Lie group to its Lie algebra
(defined in suitable representations) and 
represent the unitaries by exponentials of the
corresponding  generators,
\[
e^{\, i \int \! dt \, \bfi(t) \bQi(t)} := S(F_\bfi) \, .
\]
Now, as we have seen, $S(F_{\ddot{\bxi}_0}) = S(0) = 1$ for the second
derivatives of all
loop functions $\bxi_0$. So the time dependence of the generators
is given by 
\[
t \mapsto \bQi(t) = \bQi + t \dot{\bQi} \, .
\]
Putting $\bai := \int \! dt \, \bfi(t)$ and 
$\bbi := \int \! dt \, t \, \bfi(t)$, we therefore obtain 
\[
\int \! dt \, \bfi(t) \, \bQi(t) = 
\bai \, \bQi + \bbi \, \dot{\bQi} = \sum_{k = 1}^{3N}
(\bai_k \bQi_k + \bbi_k  \dot{\bQi}_k) \, .
\]
In view of the freedom to choose the functions $\bfi$, it 
then follows from the above 
Weyl relations for the exponentials that
\[
[\bQi_k, \dot{\bQi}_l] = i \delta_{k l} 1 \, , \quad 
[\bQi_k, \bQi_l] = [\dot{\bQi}_k, \dot{\bQi}_l] = 0 \, .
\]
Identifying $\bQi$ with the observable ``position'' and
$\dot{\bQi}$ with ``momentum'', we have thus arrived at the 
Heisenberg commutation relations. With our choice of scale
for the constant functionals, Planck's constant is put 
equal to~$1$ (atomic units). We also find that the 
time evolution of the operators is given by the solution of 
the Heisenberg equation
$ \dot \bQi(t) =i \, [H_0, \bQi(t)]$, where $H_0 = (1/2) \, \dot \bQi^2$
is the Hamiltonian corresponding to the 
underlying Lagrangean. So
we have recovered from our macroscopic point of view 
the structure of quantum mechanics. 

Interacting systems of particles can now be described by 
making use of an abstract version of the interaction 
picture. Consider a Lagrangean of the form 
\[
\Lag(\bxi, \dot{\bxi})  = (1/2) \, \dot{\bxi}^2 - V_I(\bxi) \, ,
\]
where $V_I$ describes the interaction. One first localizes 
it in time by some characteristic function $t \mapsto \chi(t)$
with arbitrary compact support.
This yields the time-dependent Lagrangean
\[
\Lag_\chi(t, \bxi, \dot{\bxi})  = (1/2) \, \dot{\bxi}^2 - \chi(t) \,  
V_I(\bxi) \, .
\]
Defining   
$\chi V_I[\bxi] := \int \! dt \, \chi(t) V_I(\bxi(t))$, 
one then introduces relative operations $S_{\Lag_\chi}$. They 
depend on functionals $F$ as in the non-interacting case which 
are given by  
\[
S_{\Lag_\chi}(F) := S(-\chi V_I)^{-1} S(F -\chi V_I) \, .
\]
Here $S$ are the operations considered before in the non-interacting case. 
Thus the operators $S_{\Lag_\chi}(F)$ are still elements of the 
non-interacting algebra $\Alg_{\Lag_0}$. Moreover, by 
some standard computation \cite{BuFr2} one finds that 
they satisfy the defining relations for the dynamical 
algebra $\Alg_{\Lag_\chi}$, where the non-interacting Lagrangean 
$\Lag_0$ in the definition given in Sec.\ 3
is replaced by $\Lag_\chi$.
So all these dynamical algebras can be accommodated in $\Alg_{\Lag_0}$.
By some more detailed analysis \cite{BuFr2}, one then 
shows that this feature prevails for the algebra $\Alg_{\Lag}$, 
which is obtained in the limit $\chi \rightarrow 1$.
To be precise, this has been established so far only for 
some large family of interaction potentials. 
Condoning this point, we conclude that there is a single 
dynamical algebra for all mechanical systems, the 
algebra~$\Alg_{\Lag_0}$. Algebras corresponding to different 
Lagrangeans merely amount to a reinterpretation of 
its elements. 

We conclude this section by noting how one can proceed from
the present abstract algebraic setting to the familiar Hilbert
space formulation in the Schr\"odinger representation,  
based on wave functions $\bxi \mapsto \psi(\bxi)$. There one 
represents $\bQi$ by multiplying the wave functions with $\bxi$ 
and $\bPi$ by taking their gradient~$-i \partial_\bxi$.  The 
operators $S(F)$, representing operations,  
are then given by time ordered exponentials,
\[
S(F) \simeq T e^{i \int \! dt \, F(\bQi + t\bPi)} \, .
\]
We also note that the familiar statistical interpretation of quantum 
mechanics in terms of expectation values of observables 
can be based on the present concept of operations. The interested reader may
consult \cite{BuFr2} for further details.

\section{Summary and outlook}
In the present letter we have established a relation between the
path integral approach to quantum physics and the framework of
dynamical algebras, established in \cite{BuFr1,BuFr2}. This new 
approach is based on a macroscopic concept of operations, describing 
perturbations of the dynamics; it dispenses with assumptions 
about their actual implementation in the quantum world. Quantum 
effects are inherited from the macroscopic arrow of time, determining
the causal structure of operations; detailed properties, such as
commutation relations, 
then follow from the given dynamics. The framework of 
dynamical algebras avoids the mathematical subtleties involved in the 
definition of path integrals; but it comprises the same physical
information. It is defined in the mathematically well established and 
convenient setting of C*-algebras. As we have shown, computations of 
integrals can be replaced by algebraic manipulations. 

Such as the path integral formalism, the framework of dynamical algebras
covers a large class of theories. It requires that classical concepts, 
such as the notion of configuration space, transformation groups, 
Lagrangeans and macroscopic causal structures provide a
physically meaningful 
starting point. With this input, it offers a complementary 
approach to quantum physics and already led to progress in the
rigorous formulation of interacting quantum field theories \cite{BuFr1}.
Within that context, it seems worthwhile to have a look at 
other longstanding problems. Examples are an algebraic version of 
Noether's theorem and the study of anomalies; the algebraic treatment 
of gauge quantum field theories and an algebraic version of the 
renormalization group. For the mathematical consolidation of 
quantum field theory in physical spacetime it would, however, be
most important to make progress in the representation theory of 
dynamical algebras which goes beyond renormalized 
perturbation theory.

\end{document}